\documentstyle[prl,aps,epsfig,floats]{revtex}

\newcommand\beq{\begin{equation}}
\newcommand\eeq{\end{equation}}
\newcommand\bea{\begin{eqnarray}}
\newcommand\eea{\end{eqnarray}}
\newcommand\non{\nonumber}
\newcommand\noi{\noindent}
\newcommand\sign{{\rm sign}}
\newcommand\bib{\bibitem}

\begin{document}

\draft

\textheight=23.8cm
\twocolumn[\hsize\textwidth\columnwidth\hsize\csname@twocolumnfalse\endcsname

\title{\Large Tunneling through two resonant levels: fixed points and 
conductances}
\author{\bf Sumathi Rao$^1$ and Diptiman Sen$^2$} 
\address{\it $^1$ Harish-Chandra Research Institute, Chhatnag Road, Jhusi,
Allahabad 211019, India \\
\it $^2$ Centre for Theoretical Studies, Indian Institute of Science, 
Bangalore 560012, India}

\date{\today}
\maketitle

\begin{abstract}
We study point contact tunneling between two leads of a Tomonaga-Luttinger 
liquid through two degenerate resonant levels in parallel. This is one of the 
simplest cases of a quantum junction problem where the Fermi statistics of the
electrons plays a non-trivial role through the Klein factors appearing in 
bosonization. Using a mapping to a `generalized Coulomb model' studied in the 
context of the dissipative Hofstadter model, we find that any asymmetry in the
tunneling amplitudes from the two leads grows at low temperatures, so that 
ultimately there is no conductance across the system. For the symmetric case,
we identify a non-trivial fixed point of this model; the conductance at that 
point is generally different from the conductance through a single resonant 
level.
\end{abstract}
\vskip .6 true cm

\vskip.5pc
]
\vskip .6 true cm

The calculation of conductances of quantum wires and dots continues to be
of major interest, as the sophistication in the fabrication of 
semiconductor heterostructures and carbon nanotubes increases.
Tunneling through barriers and quantum dots are now
amenable to sophisticated controls that allow tunneling through
specific levels in quantum dot. On the theoretical side, 
several studies of low dimensional systems have probed the effects
of strong correlations which often lead to novel features \cite{gogolin}. 

The motivation for this work is that tunneling through two resonant levels 
appears to be the simplest set-up in which the Fermi statistics of the 
electrons that tunnel through the levels play a non-trivial role and lead to 
a non-trivial fixed point. Further, using quantum dots in parallel, 
the set-up can also be experimentally realized. 

In this paper, we study a model of quantum tunneling of spinless electrons
between two leads through two parallel resonant levels which are degenerate. 
The set-up is schematically shown in Fig. 1. A simple experimental realization
of the model is shown in Fig. 2; the leads labeled $A$ and $B$ are connected to
two quantum dots labeled 1 and 2 in parallel, with both the dots having been
tuned to conduct through a single level. A magnetic flux can also be passed 
through the ring allowing for arbitrary phases in the tunneling amplitudes. 

\begin{figure}[htb]
\begin{center}
\epsfig{figure=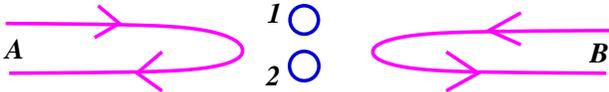,width=8.5cm}
\end{center}
\caption{Schematics of a two-lead two resonant level multiple tunneling setup.}
\end{figure}

\begin{figure}[htb]
\begin{center}
\epsfig{figure=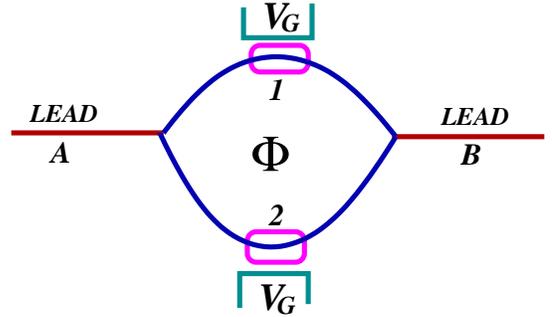,width=7.5cm}
\end{center}
\caption{Possible experimental realization of tunneling through two resonant 
levels, using two quantum dots in parallel which are tuned to resonance by 
gate voltages. Phases in the tunneling amplitudes can be introduced through
a non-zero magnetic flux through the ring.}
\end{figure}

The bosonized action for the two disconnected wires in Fig. 1 is given in
imaginary time by \cite{gogolin}
\beq
S_0 ~=~ \frac{1}{2g} ~\int d\tau ~[\int_{-\infty}^0 dx (\partial_\mu \phi_A)^2
+ \int_0^\infty dx (\partial_\mu \phi_B)^2 ]~.
\eeq
Here $\phi_a$ ($a=A,B$) is related to the electron annihilation operator
on the $a^{th}$ wire as $\psi_a \sim \eta_a e^{i\phi_a /{\sqrt 2}}$.
The Klein factors $\eta_a$ are needed to enforce Fermi statistics, 
i.e., $\{\eta_a,\eta_b\} = 2\delta_{ab}$. The $\eta_a$'s can be chosen to be
two of the Pauli matrices; hence $\eta_a^{\dagger} = \eta_a$.

The two charge states at each of the resonant levels at the origin are
represented by spin 1/2 degrees of freedom. The spin raising and
the spin lowering operators $S_i^\pm$ are the creation and
annihilation operators for an electron at the resonant level $i$.
The tunneling between the leads and the resonant levels is described by
\bea
H_t &=& t_{1A}S_1^+\eta_A e^{i\phi_A /{\sqrt 2}} ~+~ t_{1B}S_1^+\eta _B 
e^{i\phi_B /{\sqrt 2}} \non \\
&+& t_{2A}S_2^+\eta_A e^{i\phi_A /{\sqrt 2}} ~+~ t_{2B}S_2^+\eta _B 
e^{i\phi_B /{\sqrt 2}} + h.c.
\label{ht}
\eea
Here $t_{(1B,2B)}$ are the amplitudes for the electrons to tunnel to the two 
resonant levels 1 and 2 from the right lead $B$, and $t_{(1A,2A)}$ are the 
tunneling amplitudes from the left lead $A$. We have been most general here and
allowed for asymmetric tunnelings. If the tunneling was symmetric between the 
two leads, we would have $t_{1A,2A}=t_{1B,2B}$. We have also allowed for 
different tunneling amplitudes to the two levels; $t_{1A,1B}$ need not be equal
to $t_{2A,2B}$. Note that in general, $t_{ia}$ ($i=1,2$ and $a=A,B$) may be 
complex; this will take care of the phases that may be generated by a flux 
through the ring. However, for the calculations below, we will assume that 
all the $t_{ia}$ are real. 

It is easy to check that the scaling dimensions
of all the tunneling operators in Eq. (\ref{ht}) are just $1/(2g)$. So for 
$g < 1/2$, it is irrelevant and the decoupled fixed point is stable.
However, for $g > 1/2$, the decoupled fixed point is unstable; 
one would then like to find the stable fixed point(s) of the theory.

To solve the problem, we need to be able to calculate correlation functions 
involving any string of operators $e^{i\phi_a}$. In the absence of Klein 
factors, this would be trivial, since the $\phi_a$ are just free bosonic 
fields. The presence of the Klein factors gives rise to non-trivial phases. 

We first study the problem with a single resonant level to illustrate our 
method, i.e., we set $t_{2A}=t_{2B} = 0$. The partition function is given by
\beq
Z ~=~ \int {\cal D} \phi_A {\cal D} \phi_B ~e^{-S(\phi_A , \phi_B)}
\eeq
where $S = S_0 + \int d\tau dx H_t$. On expanding this, terms involving 
multiple tunnelings of the electrons will be generated. Following Ref. 
\cite{kane},
we can map the problem to a one-dimensional Coulomb gas of logarithmically 
interacting charges by classifying the different tunnelings in terms of the 
charges $q_i = m_i - n_i$ and $r_i = m_i + n_i$. Here $m_i$ and $n_i$ denote 
the charges transferred to the resonant level from the left lead and right lead
respectively. Thus, $q_i$ is the total charge transferred from the left lead 
to the right lead in any process, and $r_i$ is the change in the charge on the
resonant level in any process. The four different kinds of tunneling events can
be classified in terms of $q_i$ and $r_i$ as follows. 

\noi 1. tunnel from lead $A$ to resonant level: $r_i=1, q_i=1$.

\noi 2. tunnel from resonant level to lead $A$: $r_i=-1, q_i=-1$.

\noi 3. tunnel from lead $B$ to resonant level: $r_i=1, q_i=-1$.

\noi 4. tunnel from resonant level to lead $B$: $r_i=-1, q_i=1$.

Furthermore, events 1 and 2 carry Klein factors of $\eta_A$, while events 3 
and 4 carry a Klein factor of $\eta_B$. Let $N_k$ (where $k=1,2,3,4$) 
represent the number of each of the above events in the partition function.

Without Klein factors, the order of the tunnelings does not matter; if we also
assume that the resonant level can accommodate any number of electrons, it is
easy to see that an arbitrary term in the expansion of the partition function 
can be written as
\bea
& & (t_{1A})^{N_1+ N_2}(t_{1B})^{N_3+N_4} < {S_1^+}^{N_1} ~{S_1^-}^{N_2} ~
{S_1^+}^{N_3}~ {S_1^-}^{N_4} > \non \\
& & <e^{i\phi_A /{\sqrt 2}} \cdot \cdot ~N_1~ {\rm times}~ \times ~
e^{-i\phi_A / {\sqrt 2}} \cdot \cdot ~N_2 ~{\rm times} ~\times \non \\
& & \quad e^{i\phi_B /{\sqrt 2}} \cdot \cdot ~N_3 ~{\rm times} ~\times ~
e^{-i\phi_B /{\sqrt 2}} \cdot \cdot ~N_4 ~{\rm times}>.
\eea
This correlation function is non-zero only when $N_1=N_2$ and $N_3=N_4$; then 
the spin terms just give unity. So the full partition function is given by
\bea
& & Z ~=~ \sum_{N_1,N_3}~(t_{1A})^{2N_1} (t_{1B})^{2N_3} 
\int d\tau_1 d\tau_2 .... d\tau_N \non \\
& & \quad ~~<e^{im_1\phi_A /{\sqrt 2}} e^{im_2\phi_A /{\sqrt 2}} \cdots 
e^{in_1\phi_B /{\sqrt 2}} e^{in_2\phi_B /{\sqrt 2}} \cdots >, \non \\
& &
\eea
where $m_i =\pm 1$, $\sum_i |m_i| =2N_1$, $n_i =\pm 1$, $\sum_i |n_i|=2N_3$. 
The total number of events is $N=2N_1+2N_3$. The correlation functions 
can easily be calculated since the bosons are free. Using
\beq
<e^{im_i \phi_A (\tau_i) /{\sqrt 2}} e^{im_j\phi_A(\tau_j) /{\sqrt 2}}>
= e^{{1\over {2g}}m_im_j \ln (\tau_i-\tau_j)^2/\tau_c^2} ~,
\eeq 
where $\tau_c$ is an infra-red cutoff, one can see that the resulting partition
function is identical to that of two independent Coulomb gases, with charges 
$m_i$ and $n_i$. The partition function can be written as
\bea
Z &=& \sum_{m_i,n_i} ~(t_{1A})^{|m_i|} (t_{1B})^{|n_i|}
\int d\tau_1 d\tau_2 ... d\tau_N \non \\
&& \quad \prod_{i<j} e^{{1\over {2g}} (m_im_j +n_in_j) \ln (\tau_i-\tau_j)^2 /
\tau_c^2~} ~,
\label{partfn1}
\eea
where $N=\sum_i (|m_i|+|n_i|)$.

For the fermionic model (with Klein factors present), the ordering of the 
tunnelings becomes important since not more than one electron can sit on a 
resonant level.
(We are considering the case of spinless electrons here). This means that the 
tunnelings on to and off the resonant level have to alternate in time. In other
words, any chain of events must satisfy the following three constraints:

\noi (i) $\sum_i q_i =0$, ~(ii) $\sum_i r_i =0$, and

\noi (iii) the $r_i$ alternate in sign.

Does the presence of two Klein factors lead to non-trivial phases in the 
partition function? The answer is no; this can be understood as follows. 
First, we note that for any pair of events $i$ and $j$, the exchange of the
two Klein factors leads to a phase,
\beq
\eta_i \eta_j = \eta_j \eta_i ~e^{i(\pi/2)(q_ir_j-q_jr_i) \sign
(\tau_i-\tau_j)} ~.
\label{phase1}
\eeq
(Here $\sign (\tau) = 1$ for $\tau > 0$ and $-1$ for $\tau < 0$). Eq. 
(\ref{phase1}) can easily be checked for the events of type 1 to 4 listed 
above. For instance, 1 followed by 2 gives a phase of 1, whereas 1 followed 
by 4 gives a phase of $-1$. Eq. (\ref{phase1}) suggests that for one 
particular ordering of two events, it is useful to introduce a factor whose 
phase is equal to half the exchange phase given in that equation, namely, 
\beq
P_{ij} ~=~ e^{i(\pi/4)(q_ir_j-q_jr_i) \sign (\tau_i-\tau_j)} ~.
\eeq
One can use Eq. (\ref{phase1}) to repeatedly exchange Klein factors so as to 
eventually bring pairs of identical Klein factors next to each other; the 
product of such a pair is equal to the identity matrix since each Klein factor
is given by a Pauli matrix. In this way, one can show that the total phase 
factor for any term in the partition function can be written purely as 
a phase factor, namely,
\beq
P ~=~ \prod_{i<j} P_{ij} ~.
\eeq
Now, using the three conditions stated above, it is easy to check that $P=1$.
This is crucially dependent on condition (iii) which says that $r_i$ has to 
alternate. Hence, the phases due to the fermionic nature of the electrons
do not play a non-trivial role in this case. So the computation of the 
correlation functions can be done as for the case without the Klein factors.

To implement the constraints (i)-(iii), it is more convenient to rewrite the 
partition function in Eq. (\ref{partfn1}) terms of $r_i$ and $q_i$; we obtain
\bea
Z &=& \sum_N \sum_{q_i} (t_{1A})^{(1+q_ir_i)/2} (t_{1B})^{(1-q_ir_i)/2}
\int d\tau_1 d\tau_2 ... d\tau_N \non \\
& & \quad \prod_{i<j} e^{{1\over {4g}} (r_i r_j +q_iq_j)\ln (\tau_i-\tau_j)^2 /
\tau_c^2}~.
\label{partfn2}
\eea 
Note that we do not need the sum over $r_i$, because once
$r_1$ is fixed, the rest are fixed by the alternation rule.
This agrees with the result in Ref. \cite{kane} for the double barrier
case, which is similar to the case of the single resonant level, as expected.

The Coulomb gas partition function can then be studied using the 
renormalization group (RG) method as done in Ref. \cite{kane}. Let us first 
discuss the symmetric case $t_{1A} = t_{1B}$. For $g > 1/2$, it was shown in 
Ref. \cite{kane} that the tunneling amplitudes for a double barrier structure
(which may be expected to have the same behavior as the single resonant level
that we are considering) grow under the RG transformation, eventually leading 
to a `healing' between the leads $A$ and $B$. We will demonstrate this below.
On the other hand, for $g < 1/2$, the tunneling amplitude $t_{1A} = t_{1B}$ 
decreases under RG as mentioned earlier; the stable fixed point then is 
the `cut' wire with zero transmission between the two leads. The conductance 
is given by $ge^2 /h$ if the wire is healed, and zero if the wire is cut. 
For the asymmetric case $t_{1A} \ne t_{1B}$, the situation is somewhat 
different. One finds the wire gets healed and the conductance is $ge^2 /h$ if
$g >1$, while the wire is cut and the conductance is zero if $g<1$ \cite{kane}.

We now consider the case of two resonant levels.
The main difference from the earlier analysis is that we now
have to consider tunnelings to resonant level 1 and level 2 independently. 
We define $r_i$ as the total charge transferred to resonant level 1,
$s_i$ as the total charge transferred to resonant level 2, and $q_i$
as the charge transferred from lead $A$ to lead $B$. Then the eight 
possible tunneling events and their charge assignments are as follows.

\noi 1. tunnel from lead $A$ to level 1: $r_i=1, s_i=0, q_i=1$.

\noi 2. tunnel from level 1 to lead $A$: $r_i=-1, s_i=0, q_i=-1$.

\noi 3. tunnel from lead $B$ to level 1: $r_i=1, s_i=0, q_i=-1$.

\noi 4. tunnel from level 1 to lead $B$: $r_i=-1, s_i=0, q_i=1$.

\noi 5. tunnel from lead $A$ to level 2: $r_i=0, s_i=1, q_i=1$.

\noi 6. tunnel from level 2 to lead $A$: $r_i=0, s_i=-1, q_i=-1$.

\noi 7. tunnel from lead $B$ to level 2: $r_i=0, s_i=1, q_i=-1$.

\noi 8. tunnel from level 2 to lead $B$: $r_i=0, s_i=-1, q_i=1$.

Events 1-2 and 5-6 have Klein factors $\eta_A$, while the others have Klein 
factors $\eta_B$. Let us again denote the number of events of type $i$ as 
$N_i$. It is easy to check that the correlation function of such a set of 
events will be non-zero only if $N_1+N_5 =N_2+N_6$ and $N_3+N_7=N_4+N_8$. The 
partition function for this model can now be computed in the same way that it 
was computed for the single resonant level case. However, as we shall see 
below, this model will get mapped to a different Coulomb gas model which has 
charges $r_i$, $s_i$ and $q_i$ and non-trivial phases. 

Just as before, Fermi statistics implies that for any chain of events,

\noi (i) $\sum_i q_i=0$, (ii) $\sum_i r_i = \sum_i s_i=0$, and

\noi (iii) non-zero values of $r_i$ and $s_i$ alternate in sign.

We now define $R_i=r_i+s_i$, and note that it can only take the values $\pm 1$.
Overall charge neutrality implies that $\sum_i R_i=0$, but $R_i$ does not have
to alternate in sign (unlike $r_i$ in the earlier model with only one resonant
level). One finds that the phase can once again be written as
\beq
P_{ij} = e^{i (\pi/4)(q_iR_j-q_jR_i){\rm sign}(\tau_i-\tau_j)}~.
\eeq
However, since the $R_i$ do not have to alternate, the total phase $P$ in a 
chain of events in the partition function does not always have to be 1; it can
sometimes be $-1$. This can be seen if we consider the chain of events 1,4,6,7
which contains the string $\eta_A \eta_B \eta_A \eta_B = -1$.
Hence, this model clearly has non-trivial phases. This means that if we want 
to map it to a Coulomb gas problem, we need to worry about phases. That 
is, in the partition function, along with the logarithms which appear due to 
the contraction of the pairs of $\phi$ fields, we also need to include the 
phases which appear from the Klein factors. Note that the above phase occurs 
even in the absence of a magnetic flux through the ring, i.e., even when the
tunneling amplitudes are real. When the tunnelings are complex (due to a flux 
through the ring), we will have extra phases.

Let us write the partition function including the phases as follows:
\bea
Z &=& \sum_N \sum_{R_i,q_i} ~(\prod_i t_{ia}) ~\int d\tau_1 d\tau_2 ...
d\tau_N \non \\
&& \quad \prod_{i<j} P_{ij} ~e^{{1\over {4g}} (R_i R_j +q_iq_j) \ln (\tau_i
-\tau_j)^2 /\tau_c^2}~,
\label{partfn3}
\eea
where $t_{ia}$ can denote $t_{1A}, t_{1B}, t_{2A}$ or $t_{2B}$.
Eq. (\ref{partfn3}) looks very similar to Eq. (\ref{partfn2})
for the model with a single resonance level except for the 
alternating constraint and the presence of the phases $P_{ij}$.

Fortunately, the above problem with the phases can be mapped to a `generalized
Coulomb gas model' studied in the context of the dissipative Hofstadter model,
{\it provided that} $t_{1A} = t_{2A} = t_A$ and $t_{1B} = t_{2B} = t_B$.
This is a model of free bosons with a `magnetic field' at the
boundary, and it has been studied in detail in Ref. \cite{callan,maldacena}.

Let us introduce the model studied in Ref. \cite{callan} and show that the 
expansion of its partition function agrees with the expansion of the partition 
function for the above model with two resonant levels, under a certain
identification of the parameters. They introduced a model for the quantum 
motion of a single particle in the presence of a magnetic field, a 
periodic potential and dissipation; the model is described by the action
\bea
S &=& {1\over 2} \int d \omega ~[~ \alpha |\omega| \delta_{\mu\nu} ~
+~ \beta \omega \epsilon_{\mu\nu} ~]~ X_\mu(\omega) X_\nu(\omega) \non \\
& & + \int d\tau ~[t_A e^{i{\bf K}_1\cdot {\bf X}(\tau)} + t_B 
e^{i{\bf K}_2\cdot {\bf X}(\tau)} + h.c.] ~,
\label{action}
\eea
where $\mu,\nu =1,2$, $\epsilon_{12} = -\epsilon_{21} =1$, and $\epsilon_{11}
= \epsilon_{22} =0$. Here 
$\alpha$ and $\beta$ are related to the dissipation and the magnetic
field respectively. The potential term is defined in terms of two vectors 
${\bf K}_1 =(1,0)$ and ${\bf K}_2 = (0,1)$; these vectors define a rectangular
basis for a two-dimensional plane defined by ${\bf X} = X_1 {\bf K}_1 + X_2
{\bf K}_2$. The quadratic part of the above action leads to the propagator
\bea
D_{\mu\nu} = & & <X_\mu(\tau_1)X_{\nu}(\tau_2)> \non \\
= & & - ~{\alpha\over \alpha^2 +\beta^2} ~\delta_{\mu\nu} ~\ln 
(\tau_1-\tau_2)^2 / \tau_c^2 \non \\
& & + ~i\pi{\beta \over \alpha^2 +\beta^2} ~\epsilon_{\mu\nu}~ {\rm sign} 
(\tau_1-\tau_2) ~.
\eea
The unusual part of the above propagator is the second term or phase term, 
which exists only in the presence of the second term in the action 
Eq. (\ref{action}); this is a `magnetic field' term and it is antisymmetric 
in the indices $\mu$ and $\nu$. If we now expand the partition function in 
powers of the perturbations $t_i$, we get terms of the form 
\bea
& & (\prod_i t_i) \int d\tau_1 d\tau_2 \dots ... d\tau_n \non \\
& & <e^{i{\bf L}_1 \cdot {\bf X}(\tau_1)} <e^{i{\bf L}_2 \cdot 
{\bf X} (\tau_2)} \dots <e^{i{\bf L}_n \cdot {\bf X}(\tau_n)}> \non \\
& & = (\prod_i t_i) \int d\tau_1 d\tau_2 \dots ... d\tau_n \delta
(\sum_i {\bf L}_i) \non \\
& & \exp ~[{\alpha\over \alpha^2 +\beta^2} ~\sum_{i<j}
{\bf L}_i\cdot{\bf L}_j {\rm ln}(\tau_i-\tau_j)^2/\tau_c^2 \non \\
& & -i\pi ~{\beta\over \alpha^2 +\beta^2} ~\sum_{i<j}{\bf L}_i\times
{\bf L}_j {\rm sign}(\tau_i-\tau_j)]
\label{expand}
\eea
at the $n^{th}$ order. Here each ${\bf L}_i$ is one of the four vectors 
$\pm {\bf K}_{1,2}$.

For the eight tunneling events described above, let us associate the vector 
${\bf K}_1$ with events 1 and 5, $-{\bf K}_1$ with events 2 and 6, ${\bf K}_2$
with events 3 and 7, and $-{\bf K}_2$ with events 4 and 8. Then we find 
that for any pair of events, $R_i R_j + q_i q_j = 2 {\bf K}_i \cdot {\bf K}_j$
and $q_i R_j - q_j R_i = - 2 {\bf K}_i \times {\bf K}_j$. Hence 
the term in Eq. (\ref{expand}) matches a similar term in the partition 
function Eq. (\ref{partfn3}) of the resonant tunneling model if we equate
\bea
\frac{\alpha}{\alpha^2 + \beta^2} &=& \frac{1}{2g} ~, \non \\
\frac{\beta}{\alpha^2 + \beta^2} &=& n ~-~ \frac{1}{2} ~,
\label{albe}
\eea
where $n$ is some fixed integer. This implies that
\bea
\alpha &=& \frac{2g}{1 ~+~ (2n-1)^2 g^2} ~, \non \\
\beta &=& \frac{2(2n-1) g^2}{1 ~+~ (2n-1)^2 g^2} ~.
\eea
Different values of the integer $n$ describe different field theories for the 
$X_i$ in Eq. (\ref{action}), but they describe the same resonant level model. 
As we have seen earlier, the tunneling term is irrelevant if $g<1/2$ (i.e., 
inside the circle $\alpha^2+\beta^2 =\alpha$ in the ($\alpha ,\beta$) plane),
and is relevant for $g>1/2$ (outside the circle $\alpha^2+\beta^2 =\alpha$).
Thus we flow towards the strong tunneling limit if $g > 1/2$. 

We now begin at the 
opposite end and study the stability of the infinite tunneling (healed) limit.
For $t_A , t_B \rightarrow \infty$ in Eq. (\ref{action}), the fields $X_1$
and $X_2$ get pinned at the minima of the potential. These minima form a square
lattice with lattice spacing $2 \pi$. The fluctuations around these minima are
given by instantons which tunnel from one minimum to another \cite{callan}. 
The scaling dimension of these fluctuations are given by
the square of the lattice spacing (in units of $2\pi$) multiplied by
$\alpha$ \cite{callan,chamon}. Hence the scaling dimension is given by
\beq
\Delta_n ~=~ \frac{2g}{1 ~+~ (2n-1)^2 g^2} ~.
\eeq
We see that this is always less than 1 except when $g=1$ and $n=0$ or 1.
Thus the strong tunneling limit is always unstable, except when $g=1$
(which describes non-interacting electrons); we shall discuss this special
case later. We therefore conclude that healing by resonant tunneling through 
two levels is generically not possible.

However, let us now return to the case of resonant tunneling through only
one wire \cite{kane}, namely, $t_{2A} = t_{2B} =0$. Then there are no
phases as we saw earlier. A comparison between Eqs. (\ref{partfn2}) and 
(\ref{expand}) shows that the parameters $\alpha$ and $\beta$ satisfy
\bea
\frac{\alpha}{\alpha^2 + \beta^2} &=& \frac{1}{2g} ~, \non \\
\frac{\beta}{\alpha^2 + \beta^2} &=& n ~.
\eea
This implies that
\bea
\alpha &=& \frac{2g}{1 ~+~ 4n^2 g^2} ~, \non \\
\beta &=& \frac{4ng^2}{1 ~+~ 4n^2 g^2} ~.
\eea
In the limit of infinite tunneling, the fields $X_1$ and $X_2$ again get 
pinned at the points of a two-dimensional lattice with lattice spacing $2\pi$.
The scaling dimension of the instanton fluctuations is now given by 
\beq
\Delta_n ~=~ \frac{2g}{1 ~+~ 4n^2 g^2} ~.
\eeq
The maximum possible value of this occurs at $n=0$, when the scaling dimension
is $2g$. Then the infinite tunneling limit is stable for $g > 1/2$ and 
unstable for $g < 1/2$. This is the result in Ref. \cite{kane}, except for 
some modifications in the region $1/4 < g < 1/2$ (these arise because they 
deal with a double barrier system, not a single resonant level, and hence have 
other possibilities of back-scattering). Ref. \cite{kane} 
also shows that the conductance is given by $ge^2 /h$ if $g> 1/2$.

A better understanding of the two-resonant level model can be obtained from the
RG equations for the theory defined in Eq. (\ref{action}). Using the method 
described in Ref. \cite{maldacena}, one can derive the RG equations to third 
order in the tunneling amplitudes $t_A$ and $t_B$. Using Eqs. 
(\ref{albe}), we find that for any integer $n$,
\bea
{dt_A \over d\ln L} &=& (1 ~- ~{1 \over 2g}~) ~t_A ~-~ {t_A t_B^2 \over 4 
\pi^2} ~, \non \\
{dt_B \over d\ln L} &=& (1 ~-~ {1 \over 2g}~) ~t_B ~-~ {t_B t_A^2 \over 4 
\pi^2} ~,
\label{rg}
\eea
where $L$ denotes the length scale. For $g>1/2$, there is a stable fixed point 
for non-zero values of the $t_a$ given by $t_A^2 = t_B^2 = t_c^2$ where
\beq
t_c^2 ~=~ 4 \pi^2 ~(1- ~{1 \over 2g}~) ~.
\eeq
The location of this fixed point moves closer to zero, the closer we get to 
$g = 1/2$. So this third order RG result is trustworthy for small values 
of $t_c$ which occur close to $g = 1/2$. Since we expect the conductance of 
the system to be proportional to the square of the tunneling amplitude (if the
amplitude is weak), we find that
\beq
G ~\sim ~\frac{e^2}{h} ~ t_c^2 ~\sim ~\frac{e^2}{h} ~(~1 ~-~ {1 \over 2g} ~)~.
\label{ntcond}
\eeq
Note the non-linear dependence on $g$ at this point, similar to the non-linear 
dependence obtained in Ref. \cite{chamon}. 

Eqs. (\ref{rg}) imply that
\beq
{d(t_A^2 - t_B^2) \over d \ln L} ~=~ (~2 ~-~ \frac{1}{g} ~)~ (t_A^2 - t_B^2) ~.
\eeq
For $g > 1/2$, this implies that any asymmetry in the amplitudes grows with 
the length scale. Hence, if $t_A^2 \ne t_B^2$ to begin with, they will
become increasingly more unequal. One can then show from Eqs. (\ref{rg})
that eventually the larger amplitude will flow to infinity while the smaller
amplitude will flow to zero.

We thus see that if $g > 1/2$, a non-zero conductance through two levels in
parallel is not a stable situation. If we fine tune the tunneling amplitudes 
so that $t_{1A} = t_{2A} = t_A$ and $t_{1B} = t_{2B} = t_B$ are equal, 
then the system flows to an intermediate fixed point (IFP). But if we begin 
with generic values of $t_A$ and $t_B$ which are not equal, then one of them 
eventually grows to infinity while the other goes to zero. The conductance 
between the two leads is then zero in the long distance limit. A 
schematic RG flow diagram in the $(t_A , t_B)$ plane is shown in Fig. 3.

\begin{figure}[htb]
\begin{center}
\epsfig{figure=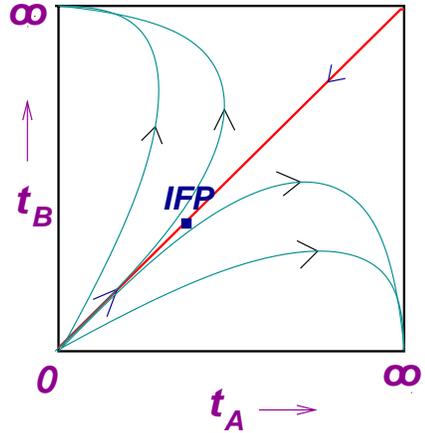,width=6cm}
\end{center}
\caption{Schematic picture of the renormalization group flow in the 
$(t_A, t_B)$ plane. An intermediate fixed point is shown on the $t_A = t_B$ 
line.}
\end{figure}

The symmetry between the two leads (channels) is not `protected' by any 
conservation law in our model. In order to access the non-trivial fixed point 
experimentally, one may therefore need to consider a different set-up in which
the two channels are labeled by spin up and spin down \cite{oreg};
this would give rise to a robust symmetry in the absence of a magnetic field.

For the non-interacting case given by $g=1$, we can compute the conductance in 
terms of the parameters $t_A$ and $t_B$ defined in the previous paragraph. We 
use the equation of motion method \cite{nayak}. We first `unfold' the left and 
right half-lines to full lines, and define the electron fields $\psi_a (x, 
\tau)$, where $a=A,B$ and $-\infty < x < \infty$. The incoming and outgoing
fermion fields are given by $\psi_a (0-, \tau)$ and $\psi_a (0+, \tau)$ 
respectively. The creation and annihilation operators for the two resonant 
levels (which lie at $x=0$) are defined as $d_i (\tau)$ and $d_i^\dagger 
(\tau)$, where $i=1,2$. In terms of these fields, the action for $g=1$ is 
given by
\bea
S &=& \int d\tau ~[ ~\sum_i d_i^\dagger \partial_\tau d_i +
\int_{-\infty}^\infty dx ~\{ ~\sum_a ~\psi_a^\dagger (\partial_\tau + i 
\partial_x) \psi_a \non \\
& & \quad \quad \quad \quad \quad + ~\delta (x) ~\sum_{a,i} ~t_a ~
(\psi_a^\dagger d_i + d_i^\dagger \psi_a) ~\} ~] ~.
\eea
The equations of motion for this system are:
\bea
[~\partial_\tau ~+~ i ~\partial_x ~] ~\psi_a (x,\tau) ~+~ \delta (x) ~t_a ~
\sum_i ~d_i (\tau) &=& 0 ~, \non \\
\partial_\tau d_i (\tau) ~+~ \sum_a ~t_a \psi_a (0, \tau) &=& 0 ~.
\label{eom}
\eea 
Here $\psi_a (0, \tau) = [\psi_a (0+, \tau) + \psi_a (0-, \tau)]/2$. We
integrate the first equation in (\ref{eom}) from $x=0-$ to $x=0+$, and 
then Fourier transform in time to obtain
\bea
i ~[~ \psi_a (0+, \omega) ~-~ \psi_a (0-, \omega) ~]~ +~ t_a ~\sum_i ~ d_i 
(\omega) &=& 0 , \non \\
\omega d_i (\omega) ~+~ \sum_a ~ t_a \psi_a (0, \omega) &=& 0 .
\eea 
We now eliminate the operators $d_i (\omega)$ and relate the outgoing fermion 
fields to the incoming fermion fields through a scattering matrix $S$, namely,
$\psi_a (0+, \omega) = \sum_b S_{ab} \psi_b (0-, \omega)$. In the limit $\omega
\rightarrow 0$ (dc conductance), we find that $S_{AA} = -S_{BB} = -(t_A^2 - 
t_B^2)/(t_A^2 + t_B^2)$, and $S_{AB} = S_{BA} = -2t_A t_B /(t_A^2 + t_B^2)$. 
The conductance is given by $e^2 /h$ times $|S_{AB}|^2$. Thus the conductance 
depends on the precise values of $t_A$ and $t_B$ if $g=1$.

The non-interacting case may be exceptional in that the conductance is a 
continuous function of the tunneling amplitudes $t_A$ and $t_B$. For the 
interacting case $g \ne 1$ (and larger than 1/2), we saw above that the 
symmetric model ($t_A = t_B$) and the asymmetric model ($t_A \ne t_B$) have 
different fixed points, and the conductance at large length scales can take 
only two different values, i.e., a finite value given in Eq. (\ref{ntcond}) 
and zero respectively.

A comparison between the symmetric one-resonant level model ($t_{1A} = t_{1B} 
\ne 0$, and $t_{2A} = t_{2B} =0$) and the symmetric two-resonant level model
($t_{1A} = t_{1B} = t_{2A} = t_{2B} \ne 0$) shows that for $g$ slightly
larger than 1/2, the conductance in the former case (where only one level 
contributes) is larger than in the latter case (where both levels contribute).
This probably happens because, in the two-resonant level model, the phase 
factors $P_{ij}$ in Eq. (\ref{partfn3}) lead 
to destructive interference between different series of tunneling events.

It is straightforward to extend the above analysis to the case in which 
$t_{1A} = t_{1B}$ and $t_{2A} = t_{2B}$ are complex. A more difficult problem 
would be to study the general two-resonant level model in which the four 
tunneling amplitudes $t_{ia}$ are all different from each other and are 
complex. (Such a generalization would allow one to examine the case in which 
there is a magnetic flux through the centre of the system as indicated in 
Fig. 2). However, it does not seem possible at present to study such a 
general model using the known Coulomb gas approach which, as mentioned above, 
requires one to assume that $t_{1A} = t_{2A}$ and $t_{1B} = t_{2B}$.

To summarize, we have shown that a Tomonaga-Luttinger liquid tunneling through
two resonant levels has a non-trivial fixed point at long distances if $g > 
1/2$ and the tunneling amplitudes from the two leads are equal. If the two 
amplitudes are not equal, then the model flows to a different fixed point in 
which one of the tunneling amplitudes and, therefore, the conductance is zero.
Thus an asymmetry between the two leads is a relevant perturbation which grows
at long distances. This behavior is similar in spirit to that of the 
one-impurity two-channel Kondo model in which an asymmetry between the 
couplings of the two channels to a spin-1/2 magnetic impurity is a relevant 
perturbation which drives the system to a fixed point that is very different 
from that of the symmetric model \cite{fabrizio}.

D. S. acknowledges financial support from a Homi Bhabha Fellowship, and from 
the Council of Scientific and Industrial Research, India through Grant No. 
03(0911)/00/EMR-II.

\end{document}